\documentclass{JHEP3}
\usepackage{graphicx,amsmath,amssymb}
\usepackage{epsfig,multicol}
\usepackage{epsf}
\usepackage{epstopdf}
\input{epsf.sty}

\usepackage{graphicx}
\usepackage{amssymb}
\usepackage{amsmath}

\def\p{\partial}

\def\half{{1\over 2}}
\def\({\left(}
\def\){\right)}
\def\[{\left[}
\def\]{\right]}

\def\e{\begin{equation}}
\def\q{\end{equation}}
\def\m{\begin{eqnarray}}
\def\n{\end{eqnarray}}

\title{Scale dependence of $f_{NL}$ in N-flation}
\author{Qing-Guo Huang \footnote{huangqg@itp.ac.cn}
\\\small{\em
Key Laboratory of Frontiers in Theoretical Physics, Institute of
Theoretical Physics, Chinese Academy of Sciences, Beijing 100190,
China} }

\abstract{
Adopting the horizon-crossing approximation, we derive the spectral index of $f_{NL}$ in general N-flation model. Axion N-flation model is taken as a typical model for generating a large $f_{NL}$ which characterizes the size of local form bispectrum. We find that its tilt $n_{f_{NL}}$ is negligibly small when all inflatons have the same potential, but a negative detectable $n_{f_{NL}}$ can be achieved in the axion N-flation with different decay constants for different inflatons. The measurement of $n_{f_{NL}}$ can be used to support or falsify the axion N-flation in the near future. 
}


\begin{document}

\section{Introduction}

In the single-component slow-roll inflation the different Fourier modes of the curvature perturbation are roughly uncorrelated and their distribution is almost Gaussian \cite{Maldacena:2002vr}. 
Using $\delta N$ formalism \cite{Starobinsky:1986fxa}, the curvature perturbation can be expanded to the non-linear orders 
\m
\zeta=\delta N=\sum_i N_{,i}\delta \phi_i+\half \sum_{i,j} N_{,ij} \delta\phi_i\delta\phi_j+{1\over 6} \sum_{i,j,k} N_{,ijk} \delta\phi_i\delta\phi_j\delta\phi_k+\ ...\ ,
\label{deltaN}
\n
where the subscript ${,i}$ denotes the derivative with respect to $\phi_i$. The above formula indicates that the curvature perturbation is determined by the evolution of the homogeneous universe corresponding both to the classical inflation trajectory and to nearby trajectories. In principle, an important difference for the multi-field model from the single-component case is that the classical trajectory is not uniquely specified by the potential, but rather has to be given as a separate piece of information. It provides an opportunity to generate a large non-Gaussianity in multi-field slow-roll inflation even at the inflationary epoch \cite{Byrnes:2008wi,Byrnes:2008zy,Kim:2010ud}, not at the end of inflation \cite{Lyth:2005qk,Sasaki:2008uc,Huang:2009vk}. See some other relevant references in \cite{Vernizzi:2006ve,Kim:2006ys,Piao:2006nm,Kim:2006te,Battefeld:2006sz,Battefeld:2007en,Langlois:2008vk,Wang:2010si,Mulryne:2010rp,Suyama:2010uj}.

On the other hand, not only the size of bispectrum $f_{NL}$ is measurable, but also its scale dependence characterized by $n_{f_{NL}}$ is possibly measured by the accurate cosmological observations, such as Planck and CMBPol. In \cite{Sefusatti:2009xu} the authors concluded that Planck and CMBPol are able to provide a 1-$\sigma$ uncertainty on the spectral index of $f_{NL}$ for local form bispectrum as follows
\m
\Delta n_{f_{NL}}\simeq 0.1 {50\over f_{NL}}{1\over \sqrt{f_{sky}}}\quad \hbox{for Planck},
\n
and
\m
\Delta n_{f_{NL}}\simeq 0.05 {50\over f_{NL}}{1\over \sqrt{f_{sky}}}\quad \hbox{for CMBPol}.
\n
Recently the scale dependence of $f_{NL}$ for inflation and curvaton model is discussed in \cite{Byrnes:2009pe,Byrnes:2010ft,Byrnes:2010xd, Huang:2010cy,Riotto:2010nh} and a large parameter space for generating a detectable $n_{f_{NL}}$ in curvaton model is illustrated in \cite{Byrnes:2010xd, Huang:2010cy}. The value of $n_{f_{NL}}$ will be an important discriminator to distinguish different models.

In this paper we focus on a class of multi-field inflation, so-called N-flation, in which there are no cross couplings among the inflaton fields or the cross couplings are negligibly small. For an example, we consider the axion N-flation with a detectably large $f_{NL}$. But  we find that $n_{f_{NL}}$ is negligibly small when all inflatons have the same potential. However, a negative $n_{f_{NL}}$ with detectably large absolute value can be obtained in the model with different decay constants for different inflatons.

Our paper is organized as follows. In Sec.~2 we derive the spectral index of $f_{NL}$ for general N-flation. We explore the observables, including $n_{f_{NL}}$, in the axion N-flation in Sec.~3. Some discussions are contained in Sec.~4.

\section{Scale dependence of $f_{NL}$ in general N-flation}

In this section we will derive the spectral index of $f_{NL}$ in general N-flation with $N_f$ uncoupled inflaton fields. The potential of inflatons is given by 
\m
V=\sum_{i=1}^{N_f} V_i(\phi_i).
\n
The dynamics of inflation is governed by 
\m
H^2&=&{1\over 3}\sum_{i=1}^{N_f} \(\half {\dot \phi}_i^2+V_i(\phi_i)\), \\
\ddot \phi_i+3H\dot \phi_i&=&-V_i',
\n
where $V_i'= \p V_i/\p \phi_i$. In this paper we work on the unit of $M_p=1$. In the slow-roll limit, the above equations are simplified to be
\m
H^2&\simeq &{V\over 3}, \label{friedm} \\
3H\dot \phi_i&\simeq& -V_i'.
\n
Similarly, we introduce some `slow-roll' parameters, such as
\m
\epsilon_i\equiv \half \(V_i'\over V_i \)^2. 
\n
From Eq.(\ref{friedm}), we find
\m
\dot H=-{1\over 6} \sum_i V_i'^2/V,
\n
and then
\m
\epsilon_H\equiv -{\dot H\over H^2}=\half {\sum_i V_i'^2\over V^2},
\n
or equivalently, 
\e
\epsilon_H=\sum_i ({V_i\over V})^2 \epsilon_i. 
\q 
Inflation happens when $\epsilon_H<1$. The multi-field slow-roll inflation can be achieved even when some inflaton fields have steep potentials because the friction term for each inflaton is contributed from all of fields.

Following \cite{Lyth:1998xn}, the number of e-folds before the end of inflation is given by  
\m
N= \sum_{i=1}^{N_f} \int_{\phi_i^{end}}^{\phi_i} {V_i\over V_i'}d\phi_i.
\label{efolds}
\n
Therefore we obtain
\m
N_{,i}={V_i\over V_i'},
\n
and 
\m
N_{,ij}=\(1-{V_iV_i''\over V_i'^2}\)\delta_{ij}.
\n
Using the $\delta N$ formalism \cite{Starobinsky:1986fxa}, the power spectrum of scalar  perturbation is given by
\m
P_\zeta=\({H\over 2\pi}\)^2 \sum_i N_{,i}^2. 
\n
A slow-roll inflation model predicts a near scale-invariant power spectrum of scalar perturbation. The deviation from exact scale invariance is described by the spectral index $n_s$ which is defined by
\m
n_s\equiv 1+{d\ln P_\zeta \over d\ln k} =1-2\epsilon_H-2{\sum_{i,j}\epsilon_{ij}N_{,i}N_{,j}\over \sum_k N_{,k}^2},
\n
where
\m
\epsilon_{ij}\equiv {V_i' \over V} {V_j'\over V}-{V_i''\over V}\delta_{ij}. 
\n
See \cite{Lyth:1998xn,Sasaki:1995aw} in detail. In this paper, we assume, without loss of generality, $\dot \phi_i<0$, so that $V_i'>0$ for $\forall \ i$. On the other hand, the gravitational wave perturbation is also generated during inflation and the amplitude of its power spectrum is determined by the inflation scale 
\m
P_T={H^2\over \pi^2/2},
\n
and then the tensor-scalar ratio $r_T$ becomes 
\m
r_T\equiv {P_T/P_\zeta}=16\left/\sum_i {1\over \epsilon_i} \right. \ .
\n
The amplitude of gravitational perturbation is also near scale-invariant and its tilt is defined by 
\m
n_T\equiv {d\ln P_T\over d\ln k}=-2\epsilon_H. 
\n
In single-field inflation, $r_T=16\epsilon_H$ which leads to a consistency relation $n_T=-r_T/8$. For $N$-field inflation, it is modified to be
\m
n_T\leq -r_T/8. 
\n
Here we used the inequality 
\m
\epsilon_H\geq 1\left/\sum_i{1\over \epsilon_i}\right. \ ,
\n
where the equality is satisfied when $\epsilon_i=\epsilon_H V/V_i$ for $\forall \ i$.

From Eq.(\ref{deltaN}), the non-Gaussianity parameter $f_{NL}$ can be written by
\m
f_{NL}={5\over 6}\sum_{i,j}{N_{,i}N_{,j}N_{,ij}\over (\sum_k N_{,k}^2)^2}.
\n
Since $N_{,ij}\propto \delta_{ij}$ for N-flation, the above formula is simplified to be
\m
f_{NL}=\sum_{i}f_{NL}^{i},
\label{sfnli}
\n
where
\m
f_{NL}^{i}={5\over 6} {N_{,i}^2N_{,ii}\over (\sum_k N_{,k}^2)^2}={5\over 6}{r_T^2\over 128}  {1\over \epsilon_i}\(1-{V_iV_i''\over V_i'^2}\). 
\label{fnli}
\n
Following \cite{Byrnes:2010ft}, the spectral index of $f_{NL}$ for N-flation is 
\m
n_{f_{NL}}={1\over f_{NL}}\sum_i f_{NL}^i (2n_{multi,i}+n_{f,i}),
\label{nfnl}
\n
where
\m
n_{multi,i}&=& 2\sum_{k,l} \epsilon_{kl}\({N_{,k}N_{,l}\over \sum_j N_{,j}^2}-\delta_{il}{N_{,k}\over N_{,i}}\), \\
n_{f,i}&=&-\sum_k {N_{,k}F_{kii}^{(2)}\over N_{,ii}},
\n
and 
\m
F_{kii}^{(2)}=-2({V_i' \over V})^2 {V_k'\over V}+{V_i''\over V}{V_k'\over V}+2 {V_i'\over V}{V_i''\over V}\delta_{ki}-{V_i'''\over V}\delta_{ki}.
\n
Considering $N_{,i}=V_i/V_i'$, we obtain
\m
n_{multi,i}&=&(1-n_s-2\epsilon_H)-4{V_i\over V}\epsilon_i \(1-{V_iV_i''\over V_i'^2}\), \label{nmulti}\\
n_{f,i}&=&4 ({V_i\over V})^2\epsilon_i-{V_i\over V}\({V_i'' \over V_i}-{V_i'''\over V_i'}\)\left/\(1-{V_iV_i''\over V_i'^2}\) \right.\ . \label{nf} 
\n
Using Eqs.(\ref{sfnli}), (\ref{fnli}), (\ref{nfnl}), (\ref{nmulti}) and (\ref{nf}), we can easily calculate the spectral index of $f_{NL}$ in general N-flation. 
The first term in (\ref{nmulti}) contribute $2(1-n_s-2\epsilon_H)$ to $n_{f_{NL}}$.
If $\epsilon_H\ll1$, $2(1-n_s-2\epsilon_H)\simeq 0.08$ for $n_s=0.96$.

In addition, we are also interested in the trispectrum in this case. See the Appendix.~\ref{trispec} in detail.

\section{Axion N-flation}

In this section we focus on axion N-flation model \cite{Kim:2010ud} in which there are $N_f$ inflaton fields and the potential of $\phi_i$ is given by
\m
V_i=\Lambda_i^4 (1-\cos\alpha_i),
\n
where $\alpha_i=2\pi \phi_i/f_i$ and $f_i$ is the i-th axion decay constant. The `slow-roll' parameters $\epsilon_i$'s are given by 
\m
\epsilon_i= {2\pi^2 \over f_i^2}{1+\cos\alpha_i\over 1-\cos\alpha_i},
\n
and 
\m
N_{,ii}=1-{V_iV_i''\over V_i'^2}={1\over 1+\cos\alpha_i}.
\n
Now the number of e-folds before the end of inflation becomes
\m
N\simeq \sum_{i=1}^{N_f} \({f_i\over 2\pi}\)^2 \ln{2\over 1+\cos\alpha_i}. 
\n
Quoting the results in \cite{Kim:2010ud}, the observables of axion N-flation are 
\m
P_\zeta&=& {H^2\over 8\pi^2}\sum_i {1\over \epsilon_i},\\
n_s&=&1- 2\epsilon_H-{8\pi^2\over 3H^2}\sum_i{\Lambda_i^4\over f_i^2}{1\over \epsilon_i}\left/ \sum_j {1\over \epsilon_j} \right. , \\
r_T&=&16\left/ \sum_i {1\over \epsilon_i} \right.  , \\
f_{NL}&=& {5\over 6} {r_T^2\over 128}\sum_i {1\over \epsilon_i}{1\over 1+\cos\alpha_i}. 
\n
Using the results in the previous section, we can calculate the spectral index of $f_{NL}$ for the axion N-flation, namely 
\m
n_{multi,i}&=&(1-n_s-2\epsilon_H)-{8\pi^2\over 3H^2}{\Lambda_i^4\over f_i^2},\\
n_{f,i}&=&4({V_i\over V})^2\epsilon_i-2{V_i\over V}\epsilon_i,
\n
and then 
\m
n_{f_{NL}}&=&2(1-n_s-2\epsilon_H)-{16\pi^2\over 3H^2} \sum_i {\Lambda_i^4 \over f_i^2} {1\over \epsilon_i}{1\over 1+\cos\alpha_i} \left/ \sum_j {1\over \epsilon_j}{1\over 1+\cos\alpha_j} \right.  \nonumber \\
&+& 2 \sum_i \(2({V_i\over V})^2-{V_i\over V}\){1\over 1+\cos\alpha_i} \left/ \sum_j {1\over \epsilon_j}{1\over 1+\cos\alpha_j} \right. .
\label{sfnl}
\n
The first and second lines in Eq.(\ref{sfnl}) are contributed by $n_{multi,i}$ and $n_{f,i}$ respectively. Since the second term in $n_{multi,i}$ is large compared to $n_{f,i}$ for $\alpha_i\simeq \pi$ which corresponds to a large $f_{NL}$, the spectral index of $f_{NL}$ is mainly determined by $n_{multi,i}$.

\subsection{Axion N-flation with the same potential for different inflatons}
\label{ans}

In this subsection we discuss the model with $\Lambda_i=\Lambda$ and $f_i=f$ for $\forall \ i$ in detail. Now the tensor-scalar ratio and $f_{NL}$ become 
\m
r_T&=&{32 \pi^2\over f^2}\left/ \sum_i{1-\cos\alpha_i\over 1+\cos\alpha_i}\right. ,\label{rtff} \\
f_{NL}&=& {10\pi^2\over 3f^2} \sum_i{1-\cos\alpha_i\over (1+\cos\alpha_j)^2} \left/\( \sum_j{1-\cos\alpha_j\over 1+\cos\alpha_j}\)^2 \right. , \label{fnlff}
\n
and the spectral indices of scalar power spectrum and $f_{NL}$ are simplified to be
\m
n_s&=&1- 2\epsilon_H-{8\pi^2 \Lambda^4\over 3H^2f^2}, \\
n_{f_{NL}}&=& {f^2\over 8\pi^2}(1-n_s-2\epsilon_H)^2  \sum_i {1\over \epsilon_i}(1-\cos\alpha_i) \left/ \sum_j {1\over \epsilon_j}{1\over 1+\cos\alpha_j} \right. \nonumber \\ 
&-&{5r_T\over 96} {1\over f_{NL}} (1-n_s-2\epsilon_H), 
\n
where
\m
{\Lambda^4\over 3H^2}= 1\left/ \sum_i(1-\cos\alpha_i)\right. ,
\n
and
\m
\epsilon_H= {2\pi^2\over f^2} \sum_i {1-\cos 2\alpha_i\over 2} \left/ \(\sum_j (1-\cos\alpha_j)\)^2 \right. .
\n 
We see that $n_{multi,i}=0$ and $n_{f_{NL}}$ is expected to be small in this simple setup. We will show the smallness of $n_{f_{NL}}$ and extend the discussions in \cite{Kim:2010ud} in the following two subsubsections.

\subsubsection{$N_*\simeq N_{tot}$}
\label{nnt}

Assuming the initial conditions for $\alpha_i$ satisfy a uniform distribution, the number of e-folds before the end of inflation is roughly given by
\m
N_{tot}\simeq {\ln 2 \over 2\pi^2}f^2 N_f. 
\n
A large enough total number of e-folds, for example not less than 60, can be achieved as long as the number of inflatons is large enough, namely $N_f\gtrsim 1.7\times 10^3/f^2$.

\begin{figure}[h]
\begin{center}
\includegraphics[width=10cm]{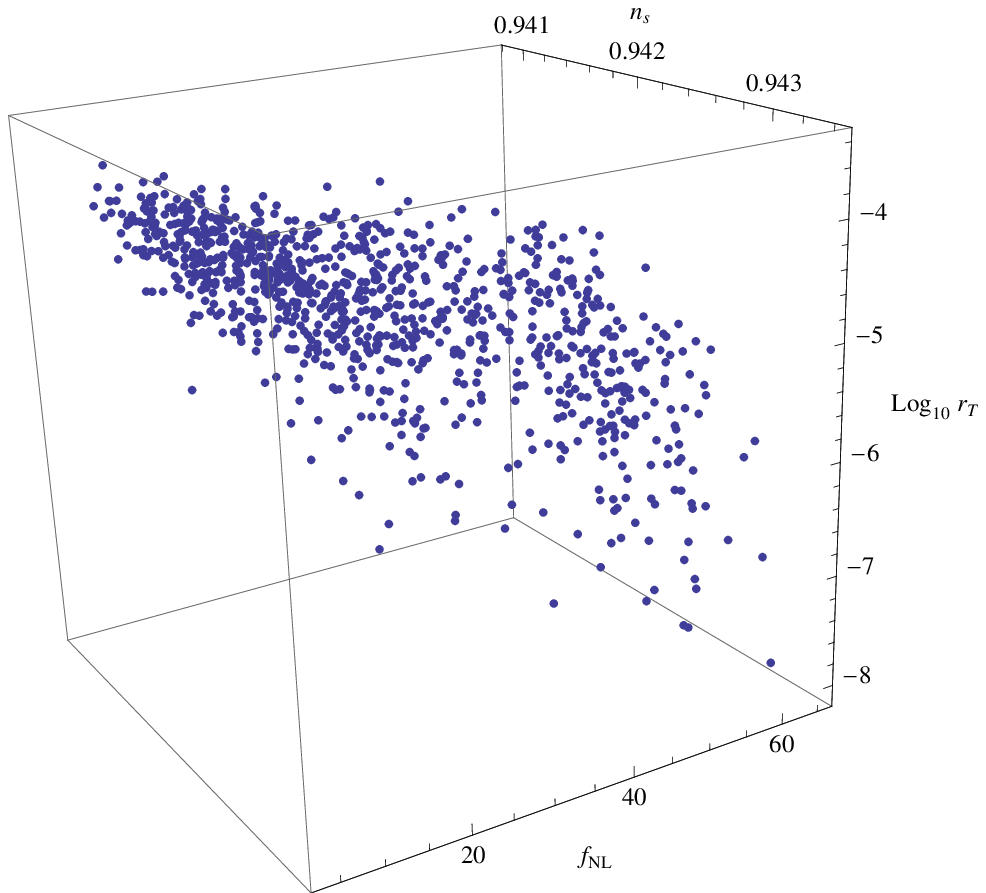}\\
\vspace{1cm}
\includegraphics[width=10cm]{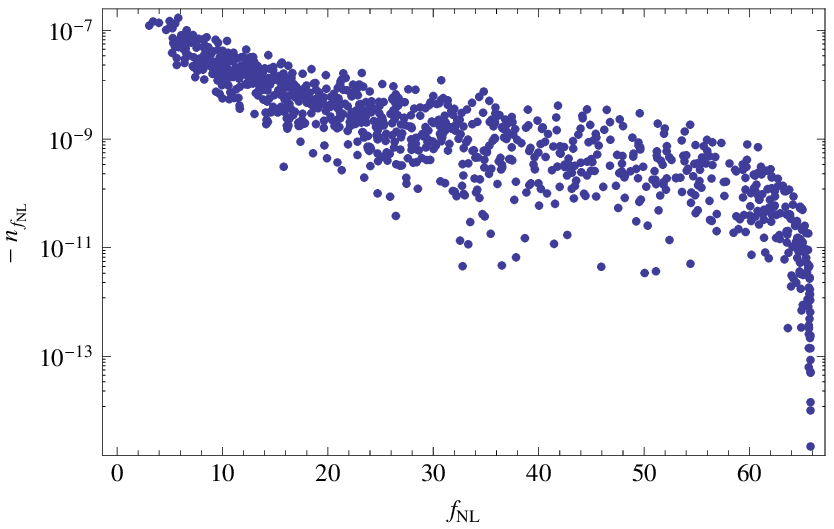}
\end{center}
\caption{The observables of axion N-flation with the same potential for different inflatons. Here $f=1/2$ and the initial conditions of $\alpha_i$ is assumed to  be the uniform distribution.}
\label{fig:random}
\end{figure}

Similar to \cite{Kim:2010ud}, we consider the case with $N_*\simeq N_{tot}$ in this subsubsection, where $N_*$ denotes the number of e-folds corresponding to the CMB scale. So we have 
\m
N_f f^2\simeq {2\pi^2N_*\over \ln 2},
\n
and $\Lambda^4/(3H^2)\simeq 1/N_f$, $\epsilon_H\simeq \pi^2/(N_f f^2)$. The spectral index of the scale power spectrum is related to $N_*$ by  
\m
n_s\simeq 1-{5\ln2 \over N_*}, 
\n
which is the same as that in \cite{Kim:2010ud}.
For $N_*\simeq 60$, $n_s\simeq 0.942$. The numerical results for $N_*\simeq N_{tot}\simeq 60$ and $f=1/2$ are illustrated in Fig.~\ref{fig:random}.
The spectral index of scalar power spectrum is nicely compatible with the estimation, and the tensor-scalar ratio is roughly ${\cal O}(10^{-5}\sim 10^{-3})$. As long as $N_f$ is large enough, from Eqs.~(\ref{rtff}) and (\ref{fnlff}), both $r_T$ and $f_{NL}$ are proportional to $1/f^2$. From the numerical results, we find that $f_{NL}$ is bounded from above, namely 
\m
f_{NL}\lesssim {5 \pi^2\over 3 f^2}. 
\label{bfnlx}
\n 
For $f\lesssim 1$, the value of $f_{NL}$ can be of order ${\cal O} (10)$ which is large enough to be detected in the near future, but its tilt $n_{f_{NL}}$ $(-{\cal O}(10^{-7})\sim -{\cal O}(10^{-14}))$ is too small to be detected in this case.

\subsubsection{A few inflatons stays around the hilltop}
\label{hilltop}

In \cite{Kim:2010ud}, the authors proposed an alternative way to achieve a large $f_{NL}$. Roughly speaking, the summation of the function which is proportional to $1/\epsilon_i$ is dominated by those fields with the smallest $\epsilon_i$, or equivalently staying around the hilltop. Suppose some number $\bar N$ of fields have roughly comparable $\epsilon_i$, of order $\bar \epsilon$ which is related to the angle $\bar \alpha$ by 
\m
\bar \epsilon\simeq {\pi^2\over f^2}(1+\cos{\bar \alpha}) \simeq {\pi^2\over 2 f^2}(\pi-{\bar \alpha})^2,
\n
with $(\pi-{\bar \alpha})\ll1$. In this case, $P_\zeta$, $r_T$, and $f_{NL}$ becomes 
\m
P_\zeta&\simeq&{H^2\over 8\pi^2} {{\bar N}\over {\bar \epsilon}}\simeq {H^2 {\bar N} f^2\over 4\pi^4 (\pi-{\bar \alpha})^2}, \label{palpha} \\
r_T&\simeq&{16{\bar \epsilon}\over {\bar N}}\simeq {8\pi^2\over {\bar N} f^2}(\pi-{\bar \alpha})^2,\\
f_{NL}&\simeq& {5\pi^2\over 3{\bar N} f^2}.  \label{fnlb}
\n
Considering ${\bar N}\geq 1$, 
\m
f_{NL}\lesssim {5\pi^2\over 3f^2},
\n
which is the same as that in Sec.~\ref{nnt}. Now the spectral index of $f_{NL}$ is given by
\m
n_{f_{NL}}\simeq -{1+\cos{\bar \alpha}\over 2} (1-n_s-2\epsilon_H),
\n
which is much smaller than $(1-n_s-2\epsilon_H)$. So the spectral index of $f_{NL}$ is undetectable in this case as well.

Because the potential become very flat around the hilltop, one may worry that the classical motions of those inflatons staying around the hilltop are not reliable. Requiring that the classical motion of inflaton ${\bar \phi}=f{\bar \alpha}/2\pi$ per Hubble time be not less than its quantum fluctuation $H/2\pi$ yields
\m
(\pi-{\bar \alpha})\gtrsim {3f H^3\over 4\pi^2 \Lambda^4}=2\xi^{-1}{H\over f}, 
\n
where $\xi=8\pi^2\Lambda^4/(3H^2f^2)=(1-n_s-2\epsilon_H)\ll 1$. \footnote{In \cite{Kim:2010ud}, the authors consider $|\phi_{hilltop}-{\bar \phi}|\gtrsim H/2\pi$, namely $(\pi-{\bar \alpha})\gtrsim H/f$, which is much looser than what we require, where $\phi_{hilltop}\equiv f/2$. } On the other hand,  from (\ref{palpha}), $(\pi-{\bar \alpha})$ can be estimated by 
\m
(\pi-{\bar \alpha})={\sqrt{\bar N} Hf\over 2\pi^2 \Delta_{\cal R}},
\n
where $\Delta_{\cal R}=\sqrt{P_\zeta}$. Therefore 
\m
\sqrt{\bar N}f^2\gtrsim {4\pi^2\Delta_{\cal R}\over \xi}. 
\n
Taking Eq.(\ref{fnlb}) into account, $f_{NL}$ is bounded from above by
\m
f_{NL}\lesssim {5\over 12}{\xi\over \Delta_{\cal R}}\left/ \sqrt{\bar N} \right. .
\n
WMAP normalization  \cite{Komatsu:2010fb} implies $P_\zeta=\Delta_{\cal R}^2=2.461\times 10^{-9}$, and then $f_{NL}\lesssim 8\times 10^3 \xi/\sqrt{\bar N}$. Since $\xi\lesssim {\cal O}(10^{-2})$, $f_{NL}\lesssim 80/\sqrt{\bar N}$. The value of $f_{NL}$ cannot be very large in this case.

\subsection{Axion N-flation with different potential for different inflatons}

In this subsection, we consider a class of axion N-flation in which $\Lambda_i=\Lambda$ for $\forall \ i$, but the decay constants can be different for different inflatons. For simplicity, we only focus on the cases with only two different decay constants. 

\subsubsection{One inflaton with different decay constant stays around the hilltop}
\label{dnnt}

In this subsubsection we consider a model with potentials
\m
V(\phi_i)&=&\Lambda^4 (1-\cos\alpha_i), \\
V(\bar\phi)&=&\Lambda^4(1-\cos{\bar \alpha}),
\n
where $\alpha_i=2\pi \phi_i/f$ for $i=1,2,...,N_f$ and $\bar\alpha=2\pi {\bar \phi}/{\bar f}$. Here $f$ is assumed to be different from $\bar f$. We consider $N_f\gg 1$ and the initial values of $\alpha_i$ satisfy a uniform distribution. \footnote{Thank C.~T.~Byrnes for the helpful discussion. } The total number of e-folds before the end of inflation is given by
\m
N_{tot}\simeq {\ln 2 \over 2\pi^2}f^2 N_f+{{\bar f}^2\over 4\pi^2}\ln{2\over 1+\cos{\bar \alpha}}.
\n
We assume $\bar\phi$ stays around the hilltop of its potential. Requiring that the classical motion of inflaton ${\bar \phi}$ is reliable leads to 
\m
(\pi-{\bar \alpha}_{ini})\gtrsim {3{\bar f} H^3\over 4\pi^2 \Lambda^4}, 
\label{dbalpha}
\n
where ${\bar \alpha}_{ini}$ denotes the initial value of ${\bar \alpha}$.

In this model, the Hubble parameter is roughly related to $\Lambda$ by $H^2\simeq N_f \Lambda^4/3$. Because the inflaton staying around the hilltop almost does not move, we also have 
\e
\epsilon_H \simeq {\ln 2\over 2}\left/\(N_{tot}-{{\bar f}^2\over 2\pi^2}\ln {2\over \pi-\bar \alpha}\) \right.\ .
\q 
The spectral index of power spectrum and $f_{NL}$ become
\m
n_s&\simeq& 1-{\ln2\over N_{tot}-{{\bar f}^2\over 2\pi^2}\ln {2\over \pi-\bar \alpha}}\[1+4 {c_r(N_f)N_f +{4\over (\pi-\bar\alpha)^2}\over c_r(N_f)N_f +(\bar f/f)^2{4\over (\pi-\bar\alpha)^2}}\], \label{nss}\\
n_{f_{NL}}&\simeq& {8 \ln2\over N_{tot}-{{\bar f}^2\over 2\pi^2}\ln {2\over \pi-\bar \alpha}}\[ {c_r(N_f)N_f +{4\over (\pi-\bar\alpha)^2}\over c_r(N_f)N_f +(\bar f/f)^2{4\over (\pi-\bar\alpha)^2}}-{d_r(N_f)N_f +{8\over (\pi-\bar\alpha)^4}\over d_r(N_f)N_f +(\bar f/f)^2{8\over (\pi-\bar\alpha)^4}}\], \nonumber \\
\n
where
\m
c_r(N)={1\over N}\sum_{i=1}^N {1-\cos\alpha_i\over 1+\cos\alpha_i}, \ \hbox{and} \
d_r(N)={1\over N}\sum_{i=1}^N {1-\cos\alpha_i\over (1+\cos\alpha_i)^2}, 
\n
which depend on the total number of points and the random choice of the series `$r$' for $\alpha_i\in[0,\pi]$. 

Similar to Sec.~\ref{nnt}, we consider that the total number of e-folds corresponds to the CMB scale, namely $N_*\simeq N_{tot}\simeq 60$.  
Considering ${\bar f}\lesssim 1$, the total number of e-fold is mainly contributed by $\phi_i$ for $i=1,2,...,N_f$. If $f={\bar f}$, $n_s\simeq 1-5\ln 2/N_*$ and $n_{f_{NL}}\simeq 0$ which are consistent with the results in Sec.~\ref{nnt}. 

\begin{figure}[h]
\begin{center}
\includegraphics[width=10cm]{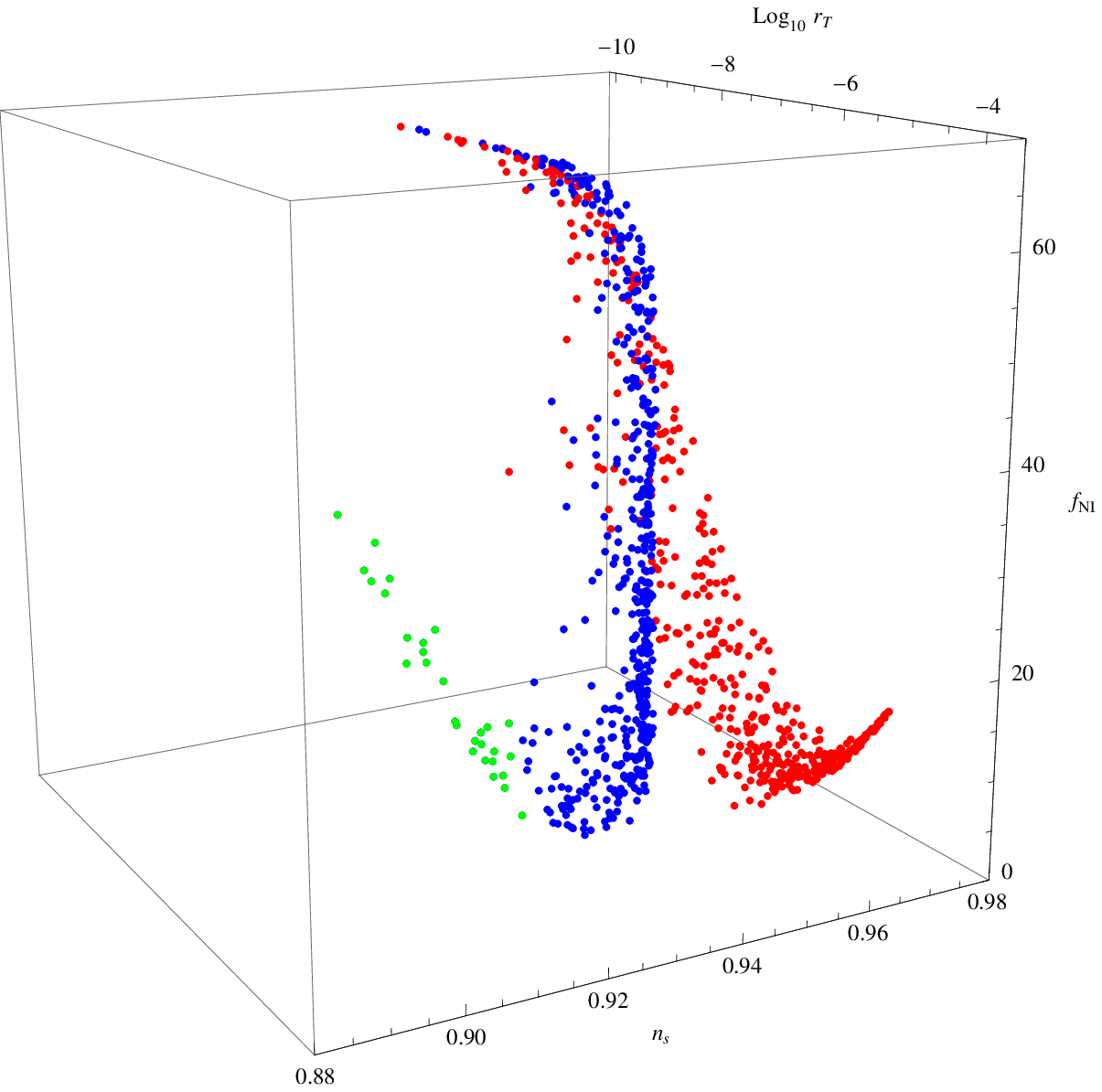}\\
\vspace{1cm}
\includegraphics[width=10cm]{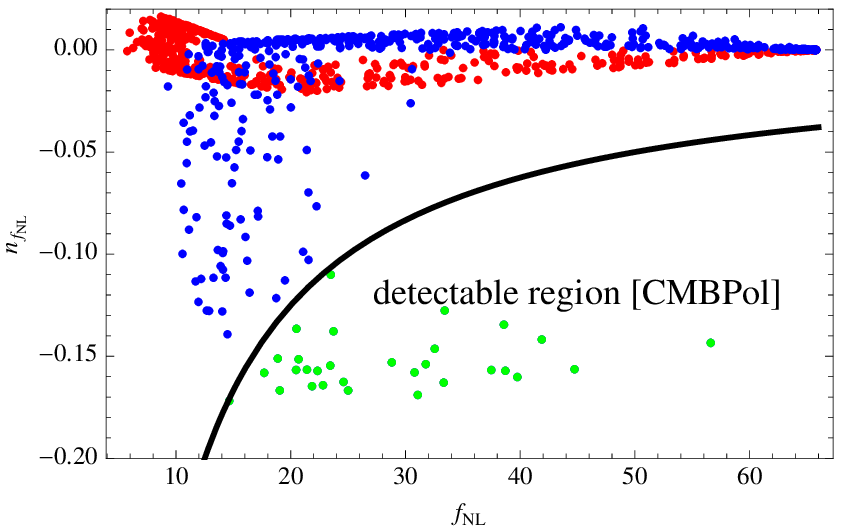}
\end{center}
\caption{The observables of axion N-flation with different decay constants. Here we adopt $f=1/2$ and $(\pi-{\bar \alpha}_{ini})/\pi=2\times 10^{-4}$. The blue and green dots correspond to ${\bar f}=1/4$ and the red dots corresponds to ${\bar f}=1$. The green dots illustrate the case with detectable scale dependence of $f_{NL}$.}
\label{fig:drandom}
\end{figure}

For $\bar f<f$, the second term in the bracket of Eq.~(\ref{nss}) is larger than 4 and then $n_s\lesssim 1-5\ln 2/N_*$. For $\bar f>f$, we find $n_s\gtrsim 1-5\ln 2/N_*$. In general, it is difficult to get the analytical estimation for the model with $f\neq \bar f$, but we can expect that the results should be quite different the model with the same decay constants. 
The numerical results are showed in Fig.~\ref{fig:drandom}. 
Here we keep $f=1/2$ and $(\pi-{\bar \alpha}_{ini})/\pi=2\times 10^{-4}$ fixed and explore the results for two choices of $\bar f$.
From Fig.~\ref{fig:drandom}, we see that a detectable negative $n_{f_{NL}}$ is possibly generated for ${\bar f}=1/4$. We also check the requirement in (\ref{dbalpha}) which is satisfied as long as $(\pi-{\bar \alpha})/\pi\gtrsim 2\times 10^{-5}$ for ${\bar f}=1/4$. Unfortunately, the spectral index $n_s\ (\lesssim 0.91)$ in the case with detectable $n_{f_{NL}}$ is small compared to WMAP 7yr data \cite{Komatsu:2010fb} $(n_s=0.968\pm 0.012)$. See the green dots in Fig.~\ref{fig:drandom}. For ${\bar f}=1$, the spectral index can fit the WMAP 7yr data nicely, but the scale dependence of $f_{NL}$ becomes negligibly small. See the red dots in Fig.~\ref{fig:drandom}. This setup indicates that a large $n_{f_{NL}}$ can be achieved in axion N-flation.

\subsubsection{A general axion N-flation with two kinds of inflatons}
\label{dnntNN}

In this subsubsection we consider a more general axion N-flation model in which there are two kinds of inflatons who have different decay constants,
\m
V_1(\phi_{1,i})&=&\Lambda^4 (1-\cos\alpha_{1,i}), \\
V_2(\phi_{2,j})&=&\Lambda^4(1-\cos\alpha_{2,j}),
\n
where $\alpha_{1,i}=2\pi \phi_{1,i}/f_1$ for $i=1,2,...,N_1$, and $\alpha_{2,j}=2\pi \phi_{2,j}/f_2$ for $j=1,2,...,N_2$. Here $f_1$ is assumed to be different from $f_2$. We also consider $N_1,\ N_2\gg 1$ and the initial values of $\alpha_{1,i}$ and $\alpha_{2,j}$ satisfy a uniform distribution respectively. 
In this model the total number of e-folds before the end of inflation is given by
\m
N_{tot}\simeq {\ln 2 \over 2\pi^2}(f_1^2 N_1+f_2^2N_2).
\n
In this model, the Hubble parameter is related to $\Lambda$ by $H^2=(N_1+N_2)\Lambda^4/3$ and the slow-roll parameter is given by
\m
\epsilon_H={\pi^2\over (N_1+N_2)^2} \(N_1/f_1^2+N_2/f_2^2\). 
\n
Similarly, we can also derive the spectral index of power spectrum and $f_{NL}$, namely  
\m
n_s&\simeq& 1-{2\pi^2\over (N_1+N_2)^2} \({N_1\over f_1^2}+{N_2\over f_2^2}\)-{8\pi^2\over N_1+N_2}{c_{r_1}(N_1)N_1 +c_{r_2}(N_2)N_2 \over c_{r_1}(N_1)f_1^2N_1 +c_{r_2}(N_2)f_2^2N_2 }, \\
n_{f_{NL}}&\simeq& {16\pi^2\over N_1+N_2} \[ {c_{r_1}(N_1)N_1 +c_{r_2}(N_2)N_2 \over c_{r_1}(N_1)f_1^2N_1 +c_{r_2}(N_2)f_2^2N_2 } - {d_{r_1}(N_1)N_1 +d_{r_2}(N_2)N_2 \over d_{r_1}(N_1)f_1^2N_1 +d_{r_2}(N_2)f_2^2N_2 } \]. \quad
\n
Again, if $f_1=f_2$, the scale dependence of $f_{NL}$ is negligibly small and our results recover those in Sec.~\ref{nnt}. 

\begin{figure}[h]
\begin{center}
\includegraphics[width=10cm]{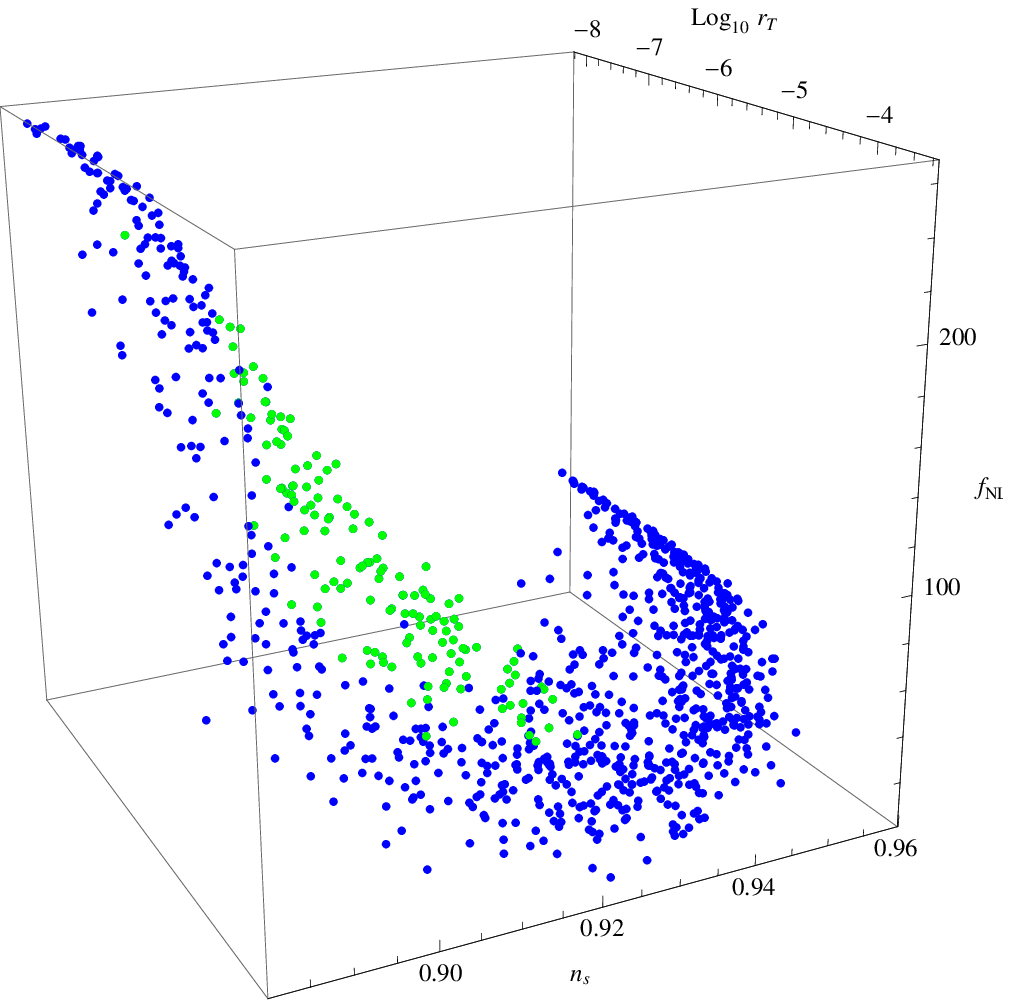}\\
\vspace{1cm}
\includegraphics[width=10cm]{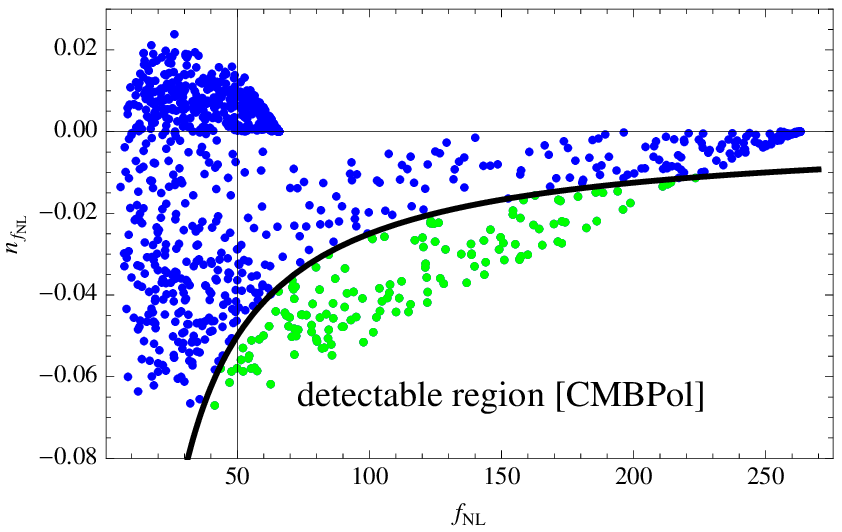}
\end{center}
\caption{The observables of a general axion N-flation with different decay constants. Here we adopt $f_1=1/2$ and $f_2=1/4$. The green dots illustrate the case with detectable scale dependence of $f_{NL}$.}
\label{fig:drandomNN}
\end{figure}

For the case with $f_1\neq f_2$, we need to adopt numerical method to explore it. Considering $N_*\simeq N_{tot}\simeq 60$, our numerical results are illustrated in Fig.~\ref{fig:drandomNN}. 
Here we adopt $f_1=1/2$, $N_1=4800$, $f_2=1/4$ and $N_2=8000$. We see that a detectable scale dependence of $f_{NL}$ can be obtained.

\section{Discussions}

In this paper we work out the spectral index of $f_{NL}$ in general N-flation model by adopting the horizon-crossing approximation. For an instance, we focus on the axion N-flation proposed in \cite{Kim:2010ud} and we find that $f_{NL}$ can be large enough to be detected in the near future. \\
$\bullet$ All inflatons have the same potential. In the case with the total number of e-folds corresponding to the CMB scale, $f_{NL}$ is roughly of order ${\cal O}(10)$ for a uniform distribution of initial angles if the decay constant $f$ is around the Planck scale. Our numerical results indicate that $f_{NL}$ is bounded from above by $5\pi^2/3f^2$, and $\tau_{NL}\simeq (2\pi/f)^{2/3}f_{NL}^{5/3}$. Another way to obtain a large $f_{NL}$ is that a few inflatons stay around the hilltop and the summations in the formula of $f_{NL}$ is mainly contributed by these inflatons. Typically $f_{NL}$ is less than one hundred in the second case. In both cases the tilt of $f_{NL}$ is negligibly small. The smallness of $n_{f_{NL}}$ in these two case comes from the accident cancellation in $n_{multi,i}$ because all inflatons have the same potential. \\
$\bullet$ Different inflatons have different decay constants. In this kind of model, the accident cancellation in $n_{multi,i}$ does not happen and a negative detectable $n_{f_{NL}}$ can be obtained. It implies that a scale independent $f_{NL}$ is not generic prediction of N-flation. A more general axion N-flation is left to be studied in the future.

In addition, we also study the trispectrum for N-flation in the Appendix.~\ref{trispec}. We find that there is a universal relation between $g_{NL}$ and $\tau_{NL}$, namely $g_{NL}\simeq 25 \tau_{NL}/27$, for the cases with large $f_{NL}$. But the relation between $f_{NL}$ and $\tau_{NL}$ depends on detail of the model.

If there is only one inflaton field, the axion N-flation becomes natural inflation. The slow-roll parameters are roughly given by $M_p^2/f^2$ and then the slow-roll conditions are satisfied if $f>M_p$. However such a large decay constant cannot be achieved in some fundamental theories, such as string theory \cite{ArkaniHamed:2003wu,Banks:2003sx}. In \cite{Dimopoulos:2005ac} the N-flation is supposed to realize a slow-roll inflation even when $f<M_p$ by calling multi inflaton fields. On the other hand, in \cite{Huang:2007st} we proposed a conjecture that the effective gravity scale should be $\Lambda_G=M_p/\sqrt{N_f}$ for the case with multi scalar fields, not $M_p$. If the decay constant of axion field is required to be smaller than $\Lambda_G$, namely $N_f f^2<1$, the total number of e-folds is less than one and the axion N-flation fails.

\vspace{1.0cm}

\noindent {\bf Acknowledgments}

\vspace{.5cm}

We would like to thank S.~A.~Kim and A.~R.~Liddle for useful discussions. 
QGH is supported by the project of Knowledge Innovation
Program of Chinese Academy of Science and a grant from NSFC.

\vspace{1.5cm}

\appendix

\section{The trispectrum in N-flation}
\label{trispec}

From Eq. (\ref{deltaN}), the size of trispectrum is characterized by two parameters $\tau_{NL}$ and $g_{NL}$ which are defined by 
\m
\tau_{NL}&=&\sum_{i,j,k} N_{,ij}N_{,ik}N_{,j}N_{,k}\left/ (\sum_l N_{,l}^2)^3 \right. ,\\
g_{NL}&=& {25\over 54} \sum_{i,j,k} N_{,ijk}N_{,i}N_{,j}N_{,k}\left/ (\sum_l N_{,l}^2)^3 \right. .
\n
For N-flation, the above two equations are simplified to be
\m
\tau_{NL}&=&\sum_i N_{,ii}^2 N_{,i}^2 \left/ (\sum_l N_{,l}^2)^3 \right. ,\\
g_{NL}&=& {25\over 54} \sum_i N_{,iii}N_{,i}^3 \left/ (\sum_l N_{,l}^2)^3 \right. ,
\n
where
\m
N_{,iii}=2{V_i V_i''^2\over V_i'^3}-{V_i''\over V_i'}-{V_iV_i'''\over V_i'^2}. 
\n
We remind the readers that the value of $\tau_{NL}$ is bounded from below by $({6\over 5}f_{NL})^2$.  
For the axion N-flation, we have
\m
N_{,iii}={2\pi\over f_i} {\sin\alpha_i \over (1+\cos\alpha_i)^2}.
\n
Therefore 
\m
\tau_{NL}&=&\sum_i {f_i^2\over 4\pi^2} {1-\cos\alpha_i\over (1+\cos\alpha_i)^3} \left/ \left(\sum_j {f_j^2\over 4\pi^2} {1-\cos\alpha_j\over 1+\cos\alpha_j}\right)^3 \right. ,\\
g_{NL}&=&{25\over 54} \sum_i {f_i^2\over 4\pi^2} {(1-\cos\alpha_i)^2 \over (1+\cos\alpha_i)^3} \left/ \left(\sum_j {f_j^2\over 4\pi^2} {1-\cos\alpha_j\over 1+\cos\alpha_j}\right)^3 \right. .
\n

\begin{figure}[h]
\begin{center}
\includegraphics[width=10cm]{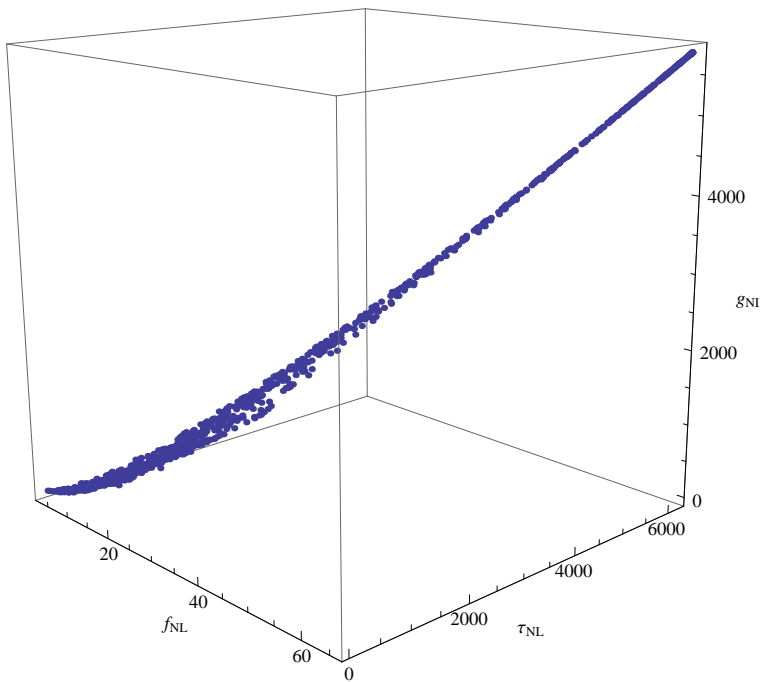}
\end{center}
\caption{The non-Gaussianity parameters of axion N-flation for the uniform distribution of $\alpha_i$ when $f=1/2$. Here we consider the case with $N_*\simeq N_{tot}\simeq 60$. }
\label{fig:gtfnl}
\end{figure}

For the case with $f_i=f$ for $\forall \ i$, both $\tau_{NL}$ and $g_{NL}$ are proportional to $1/f^4$. Similar to Sec.~\ref{nnt}, we consider $N_*\simeq N_{tot}\simeq 60$ and the numerical results for $f=1/2$ are showed in Fig.~\ref{fig:gtfnl}.
Fitting the numerical results, we find that $\tau_{NL}$ and $g_{NL}$ are roughly related to $f_{NL}$ and $f$ by
\m
\tau_{NL}\simeq \({2\pi\over f}\)^{2/3} f_{NL}^{5/3}, \quad 
g_{NL}\simeq {25\over 27} \tau_{NL}. 
\label{fft}
\n
Roughly speaking, $f_{NL}\sim 1/f^2$ and then $g_{NL}\simeq {25\over 27}\tau_{NL}\sim 1/f^4$ which is consistent with our estimation. On the other hand, combining (\ref{fft}) and $\tau_{NL}\geq ({6\over 5}f_{NL})^2$, we get $f_{NL}\lesssim 4\pi^2/3f^2$ which is roughly the same as our results in Sec.~\ref{nnt}.

The non-Gaussianity parameters of trispectrum for the case in which a few inflatons stay around the hilltop were worked out in \cite{Kim:2010ud}: $\tau_{NL}=({6\over 5}f_{NL})^2$ and $g_{NL}= {25\over 27} \tau_{NL}$, which are different from the previous case unless $f_{NL}\simeq 4\pi^2/3f^2$.

\begin{figure}[h]
\begin{center}
\includegraphics[width=10cm]{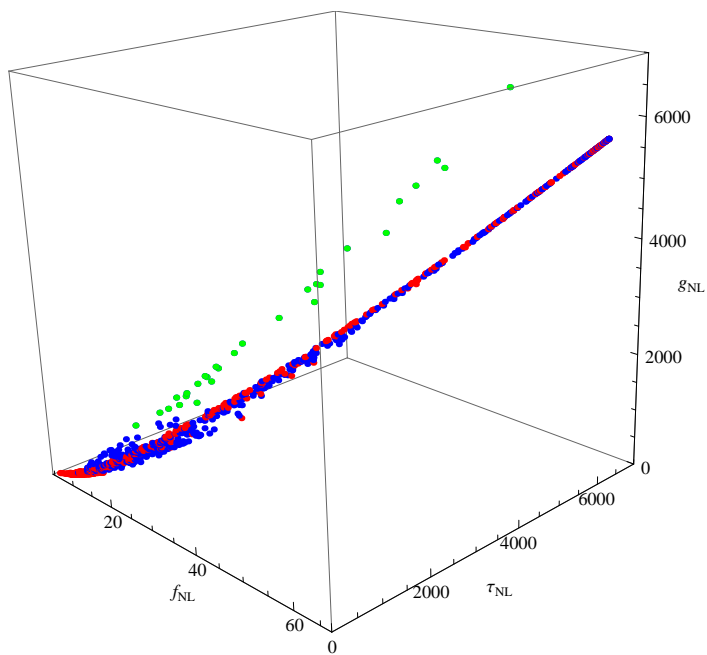}
\end{center}
\caption{The non-Gaussianity parameters in the axion N-flation where $N_f$ inflatons have the same decay constant $f$, and the decay constant of another inflaton is $\bar f$. Here we consider the case with $N_*\simeq N_{tot}\simeq 60$ and adopt $f=1/2$ and $(\pi-{\bar \alpha}_{ini})/\pi=2\times 10^{-4}$. The blue and green dots correspond to ${\bar f}=1/4$ and the red dots corresponds to ${\bar f}=1$. The green dots illustrate the case with detectable scale dependence of $f_{NL}$. }
\label{fig:dgtfnl}
\end{figure}

The non-Gaussianity parameters for the model in Sec.~\ref{dnnt} are showed in Fig.~\ref{fig:dgtfnl}.
Fitting the green dots corresponding to a detectable scale dependence of $f_{NL}$, we find 
\m
\tau_{NL}\simeq 7.78 f_{NL}^{1.77},\quad 
g_{NL}\simeq {25\over 27}\tau_{NL}. 
\n
Fitting the blue and red dots respectively, we reach the same results  
\m
\tau_{NL}\simeq&8.28 f_{NL}^{1.59},\quad
g_{NL}\simeq {25\over 27}\tau_{NL}. 
\n
Here the relations between $\tau_{NL}$ and $f_{NL}$ depend on the choice of the parameters, such as $f$, $\bar f$ and $\bar \alpha$, in the model. 

\begin{figure}[h]
\begin{center}
\includegraphics[width=10cm]{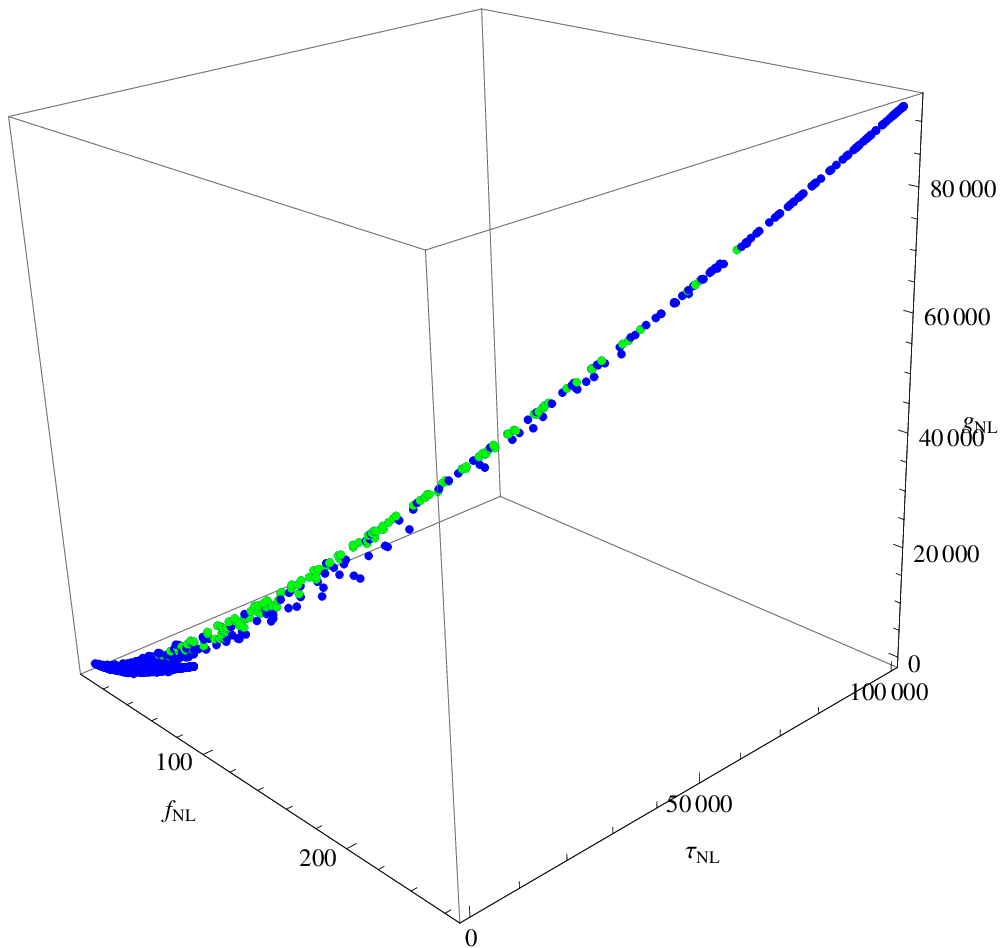}
\end{center}
\caption{The non-Gaussianity parameters in the general axion N-flation with two kinds of inflatons. Here we consider the case with $N_*\simeq N_{tot}\simeq 60$. The green dots illustrate the case with detectable scale dependence of $f_{NL}$. }
\label{fig:dgtfnlNN}
\end{figure}

Similarly we also investigate the non-Gaussianity parameters for the model in Sec.~\ref{dnntNN}. See the numerical results in Fig.~\ref{fig:dgtfnlNN}.
Fitting the blue and green dots respectively, we find  
\m
\tau_{NL}\simeq 7.16 f_{NL}^{1.72},\quad 
g_{NL}\simeq {25\over 27}\tau_{NL}. 
\n

To summarize, there is a universal relation between $g_{NL}$ and $\tau_{NL}$, namely $g_{NL}\simeq {25\over 27}\tau_{NL}$. The value of $g_{NL}$ is the same order of $\tau_{NL}$.  But the relations between $\tau_{NL}$ and $f_{NL}$ are different for different cases.


\newpage

\begin{thebibliography}{99}



\bibitem{Maldacena:2002vr}
  J.~M.~Maldacena,
  ``Non-Gaussian features of primordial fluctuations in single field
  inflationary models,''
  JHEP {\bf 0305}, 013 (2003)
  [arXiv:astro-ph/0210603].



\bibitem{Starobinsky:1986fxa}
  A.~A.~Starobinsky,
  ``Multicomponent de Sitter (Inflationary) Stages and the Generation of
  Perturbations,''
  JETP Lett.\  {\bf 42} (1985) 152;\\  
  M.~Sasaki and E.~D.~Stewart,
  ``A General Analytic Formula For The Spectral Index Of The Density
  Perturbations Produced During Inflation,''
  Prog.\ Theor.\ Phys.\  {\bf 95}, 71 (1996)
  [arXiv:astro-ph/9507001];\\ 
  M.~Sasaki and T.~Tanaka,
  ``Super-horizon scale dynamics of multi-scalar inflation,''
  Prog.\ Theor.\ Phys.\  {\bf 99}, 763 (1998)
  [arXiv:gr-qc/9801017];\\
    D.~H.~Lyth, K.~A.~Malik and M.~Sasaki,
  ``A general proof of the conservation of the curvature perturbation,''
  JCAP {\bf 0505}, 004 (2005)
  [arXiv:astro-ph/0411220];\\  
  D.~H.~Lyth and Y.~Rodriguez,
  ``The inflationary prediction for primordial non-gaussianity,''
  Phys.\ Rev.\ Lett.\  {\bf 95}, 121302 (2005)
  [arXiv:astro-ph/0504045].


\bibitem{Byrnes:2008wi}
  C.~T.~Byrnes, K.~Y.~Choi and L.~M.~H.~Hall,
  ``Conditions for large non-Gaussianity in two-field slow-roll inflation,''
  JCAP {\bf 0810}, 008 (2008)
  [arXiv:0807.1101 [astro-ph]].


\bibitem{Byrnes:2008zy}
  C.~T.~Byrnes, K.~Y.~Choi and L.~M.~H.~Hall,
  ``Large non-Gaussianity from two-component hybrid inflation,''
  JCAP {\bf 0902}, 017 (2009)
  [arXiv:0812.0807 [astro-ph]].

\bibitem{Kim:2010ud}
  S.~A.~Kim, A.~R.~Liddle and D.~Seery,
  ``Non-gaussianity in axion Nflation models,''
  arXiv:1005.4410 [astro-ph.CO].


\bibitem{Lyth:2005qk}
  D.~H.~Lyth,
  ``Generating the curvature perturbation at the end of inflation,''
  JCAP {\bf 0511}, 006 (2005)
  [arXiv:astro-ph/0510443].
  
\bibitem{Sasaki:2008uc}
  M.~Sasaki,
  ``Multi-brid inflation and non-Gaussianity,''
  Prog.\ Theor.\ Phys.\  {\bf 120}, 159 (2008)
  [arXiv:0805.0974 [astro-ph]].
  
\bibitem{Huang:2009vk}
  Q.~G.~Huang,
  ``A geometric description of the non-Gaussianity generated at the end of
  multi-field inflation,''
  JCAP {\bf 0906}, 035 (2009)
  [arXiv:0904.2649 [hep-th]].



\bibitem{Vernizzi:2006ve}
  F.~Vernizzi and D.~Wands,
  ``Non-Gaussianities in two-field inflation,''
  JCAP {\bf 0605}, 019 (2006)
  [arXiv:astro-ph/0603799].

\bibitem{Kim:2006ys}
  S.~A.~Kim and A.~R.~Liddle,
  ``Nflation: Multi-field inflationary dynamics and perturbations,''
  Phys.\ Rev.\  D {\bf 74}, 023513 (2006)
  [arXiv:astro-ph/0605604].

\bibitem{Piao:2006nm}
  Y.~S.~Piao,
  ``On perturbation spectra of N-flation,''
  Phys.\ Rev.\  D {\bf 74}, 047302 (2006)
  [arXiv:gr-qc/0606034];\\
  R.~G.~Cai, B.~Hu and Y.~S.~Piao,
 ``Entropy Perturbations in N-flation Model,''
Phys.\ Rev.\ D {\bf 80}, 123505 (2009) 
[arXiv:0909.5251 [astro-ph.CO]].

\bibitem{Kim:2006te}
  S.~A.~Kim and A.~R.~Liddle,
  ``Nflation: Non-gaussianity in the horizon-crossing approximation,''
  Phys.\ Rev.\  D {\bf 74}, 063522 (2006)
  [arXiv:astro-ph/0608186].

\bibitem{Battefeld:2006sz}
  T.~Battefeld and R.~Easther,
  ``Non-gaussianities in multi-field inflation,''
  JCAP {\bf 0703}, 020 (2007)
  [arXiv:astro-ph/0610296].


\bibitem{Battefeld:2007en}
  D.~Battefeld and T.~Battefeld,
  ``Non-Gaussianities in N-flation,''
  JCAP {\bf 0705}, 012 (2007)
  [arXiv:hep-th/0703012].


\bibitem{Langlois:2008vk}
  D.~Langlois, F.~Vernizzi and D.~Wands,
  ``Non-linear isocurvature perturbations and non-Gaussianities,''
  JCAP {\bf 0812}, 004 (2008)
  [arXiv:0809.4646 [astro-ph]].

\bibitem{Mulryne:2010rp}
  D.~J.~Mulryne, D.~Seery and D.~Wesley,
  ``Moment transport equations for the primordial curvature perturbation,''
  arXiv:1008.3159 [astro-ph.CO].

\bibitem{Wang:2010si}
  T.~Wang,
  ``Note on Non-Gaussianities in Two-field Inflation,''
  arXiv:1008.3198 [astro-ph.CO].

\bibitem{Suyama:2010uj}
  T.~Suyama, T.~Takahashi, M.~Yamaguchi and S.~Yokoyama,
  ``On Classification of Models of Large Local-Type Non-Gaussianity,''
  arXiv:1009.1979 [astro-ph.CO].


\bibitem{Sefusatti:2009xu}
  E.~Sefusatti, M.~Liguori, A.~P.~S.~Yadav, M.~G.~Jackson and E.~Pajer,
  ``Constraining Running Non-Gaussianity,''
  JCAP {\bf 0912}, 022 (2009)
  [arXiv:0906.0232 [astro-ph.CO]].



\bibitem{Byrnes:2009pe}
  C.~T.~Byrnes, S.~Nurmi, G.~Tasinato and D.~Wands,
  ``Scale dependence of local $f_{NL}$,''
  JCAP {\bf 1002}, 034 (2010)
  [arXiv:0911.2780 [astro-ph.CO]].


\bibitem{Byrnes:2010ft}
  C.~T.~Byrnes, M.~Gerstenlauer, S.~Nurmi, G.~Tasinato and D.~Wands,
  ``Scale-dependent non-Gaussianity probes inflationary physics,''
  arXiv:1007.4277 [astro-ph.CO].
  
\bibitem{Byrnes:2010xd}
  C.~T.~Byrnes, K.~Enqvist and T.~Takahashi,
  ``Scale-dependence of Non-Gaussianity in the Curvaton Model,''
  arXiv:1007.5148 [astro-ph.CO].

\bibitem{Huang:2010cy}
  Q.~G.~Huang,
  ``Negative spectral index of $f_{NL}$ in the axion-type curvaton model,''
  arXiv:1008.2641 [astro-ph.CO].

\bibitem{Riotto:2010nh}
  A.~Riotto and M.~S.~Sloth,
  ``Strongly Scale-dependent Non-Gaussianity,''
  arXiv:1009.3020 [astro-ph.CO].


\bibitem{Lyth:1998xn}
  D.~H.~Lyth and A.~Riotto,
  ``Particle physics models of inflation and the cosmological density
  perturbation,''
  Phys.\ Rept.\  {\bf 314}, 1 (1999)
  [arXiv:hep-ph/9807278].  
  
\bibitem{Sasaki:1995aw}
  M.~Sasaki and E.~D.~Stewart,
  ``A General Analytic Formula For The Spectral Index Of The Density
  Perturbations Produced During Inflation,''
  Prog.\ Theor.\ Phys.\  {\bf 95}, 71 (1996)
  [arXiv:astro-ph/9507001].

\bibitem{Komatsu:2010fb}
  E.~Komatsu {\it et al.},
  ``Seven-Year Wilkinson Microwave Anisotropy Probe (WMAP) Observations:
  Cosmological Interpretation,''
  arXiv:1001.4538 [astro-ph.CO].


\bibitem{ArkaniHamed:2003wu}
  N.~Arkani-Hamed, H.~C.~Cheng, P.~Creminelli and L.~Randall,
  ``Extranatural inflation,''
  Phys.\ Rev.\ Lett.\  {\bf 90}, 221302 (2003)
  [arXiv:hep-th/0301218].  

\bibitem{Banks:2003sx}
  T.~Banks, M.~Dine, P.~J.~Fox and E.~Gorbatov,
  ``On the possibility of large axion decay constants,''
  JCAP {\bf 0306}, 001 (2003)
  [arXiv:hep-th/0303252].


\bibitem{Dimopoulos:2005ac}
  S.~Dimopoulos, S.~Kachru, J.~McGreevy and J.~G.~Wacker,
  ``N-flation,''
  JCAP {\bf 0808}, 003 (2008)
  [arXiv:hep-th/0507205].


\bibitem{Huang:2007st}
  Q.~G.~Huang,
  ``Weak Gravity Conjecture for the Effective Field Theories with N Species,''
  Phys.\ Rev.\  D {\bf 77}, 105029 (2008)
  [arXiv:0712.2859 [hep-th]].
  






\end{thebibliography}
\end{document}